\documentstyle[epsf,rotate,aps,eqsecnum]{revtex}
\begin{document}
\vskip 2.0cm
\title{\huge {COMPOSED ENSEMBLES \\
 OF RANDOM UNITARY MATRICES}}
\vskip 2.0cm
\author{\large{Marcin Po\'zniak$^1$\footnote{deceased},
 Karol \.Zyczkowski$^2$   and Marek Ku\'s$^3$}}
\vskip 2.0cm
\address{
$^1$Instytut Matematyki, Uniwersytet Jagiello\'nski, ul. Reymonta 4,
30--057, Krak\'ow, Poland\\
$^2$Instytut Fizyki im. M. Smoluchowskiego, Uniwersytet Jagiello\'nski,\\
ul. Reymonta 4, 30--057, Krak\'ow, Poland\\
$^3$Centrum Fizyki Teoretycznej, Polska Akademia Nauk, \\
al. Lotnik{\'o}w 32/46, 02-668, Warszawa, Poland
   }
\vskip 2.0cm
\date{June 6, 1997 ~ ~ and ~ ~ October 1998}
\maketitle
\vskip 3.0cm
\centerline{\bf{\small {PREPRINT IMUJ 1997/19}}}
\vskip 3.0cm
\centerline{\bf{\small {J. Phys. A31, 1059-1071 (1998)  }}}
\vskip 3.0cm
\centerline{
{\small e-mail:
$^2$karol@chaos.if.uj.edu.pl \quad $^3$marek@ifpan.edu.pl}}

\vskip 1.0cm
\begin{abstract}
Composed ensembles of random unitary matrices are defined via products
of matrices, each pertaining to a given canonical circular ensemble of
Dyson. We investigate statistical properties of spectra of some composed
ensembles  and demonstrate their physical relevance. We discuss also the
methods of generating random matrices distributed according to invariant
Haar measure on the orthogonal and unitary group.

\end{abstract}

\newpage

\section{Introduction}
\label{s:introduction}

Random unitary matrices are often used to describe the process of
chaotic scattering \cite{bs88,jp95}, conductance in mesoscopic
systems\cite{cond} and statistics of quantum, periodically driven
systems (see \cite{haake} and references therein). They may be defined
by circular ensembles of unitary matrices, first considered by Dyson
\cite{Dys1}. He defined circular orthogonal, unitary or symplectic
ensembles (COE, CUE and CSE), which display different transformation
properties \cite{mehta}.  Distribution of matrix elements and their
correlations are known for these canonical ensembles
\cite{pm83,mps85,m90,dsf91}.

Our investigations are motivated by many successful applications of the
Random Matrix Theory to problems of quantum chaos, i.e.\ to the
description of quantum properties of systems chaotic in the classical
limit.  Random matrices of three canonical circular ensembles appear to
provide quantitatively verifiable predictions concerning statistical
properties of quasi-energy spectra, transition amplitudes etc.\ for
quantum chaotic systems \cite{haake}. For systems without generalized
time--reversal symmetry one should use CUE, while COE consisting of
unitary symmetric matrices corresponds to the time reversal invariant
systems (with integer spin). The so-called circular Poissonian ensemble
(CPE) of diagonal unitary matrices with independent unimodular
eigenvalues has also found applications for certain classically
integrable systems.

In this paper we shall study statistical properties of composed
ensembles defined by products of unitary matrices, each drawn with a
given probability distribution. Products of matrices come into attention
in a natural way when we consider the evolution of kicked systems.
Unitary propagators transporting wave functions of such systems over one
period of the kicking perturbation are products of ``free'' evolution
propagators and unitary transformations corresponding to instantaneous
kicks. Moreover, products of two unitary matrices appear also in the
theory of chaotic scattering \cite{smil,gopar}.

The paper is organized as follows. In section 2 we briefly recall the
necessary definitions and introduce notation. Section 3 contains results
concerning spectral properties of composed ensembles of random unitary
matrices. The paper is completed by concluding remarks. In appendices we
review methods of generating random matrices according to the invariant
Haar measures on the orthogonal and unitary group.

\section{Canonical ensembles of unitary matrices}
\label{s:canon}

Circular ensembles of matrices where defined by Dyson \cite{Dys1} as the
subsets of the set of unitary matrices. Uniqueness of the ensembles is
imposed by introducing measures invariant under appropriate groups of
transformations \cite{Hu63}. Specifically the circular unitary ensemble (CUE)
consists of all unitary matrices with the (normalized) Haar measure
$\mu_U$ on the unitary group $U_N$.  The circular orthogonal ensemble (COE)
is defined on the set $S_N$ of all symmetric unitary matrices
$S=S^T=(S^{\dag})^{-1}$ by the property of being invariant under all
transformations by an arbitrary unitary matrix $W$,
\begin{equation}
S\rightarrow W^TSW,
\label{wtsw}
\end{equation}
where ${}^T$ denotes the transposition. The normalized measure on COE will
be denoted by $\mu_S$.

Eigenvalues of an $N\times N$ unitary matrix lie on the unit circle,
$\lambda_i=\exp(i\varphi_i)$; $0\le\varphi_i\le 2\pi$, $i=1,\dots,N$.
The joint probability distribution (JPD) of eigenvalues for each ensemble was
given by Dyson \cite{Dys1}
\begin{equation}
P_{\beta}(\varphi_1,\dots,\varphi_N) = C_{\beta} \prod_{i<j}
  |e^{i\varphi_i} - e^{i\varphi_j}|^{\beta},
\label{joint}
\end{equation}
where $C_{\beta}$ is a normalization constant and $\beta$ equals to $1$
and $2$ for COE and CUE, respectively. This number is sometimes called
repulsion parameter, since it determines the behaviour of levels
spacings as $P(s)\sim s^{\beta}$ for small $s$ \cite{mehta}.

The above formula with $\beta=0$ describes spectra of circular
Poissonian ensemble (CPE) of diagonal unitary matrices with $N$
independent phases drawn with uniform distribution in $[0,2\pi)$. The
set of diagonal matrices will be denoted by $D_N$ and the normalized
measure on CPE, (which is simply the product measure of $N$ measures on
the unit circle,) by $\mu_D$. For further consideration we will also
need an ensemble of orthogonal matrices with the probability density
$\mu_O$ defined by the (normalized) Haar measure on the orthogonal group
in $N$ dimension.
We shall call this ensemble as Haar orthogonal ensemble (HOE). It is
invariant with respect to all transformations $O_1\rightarrow
O_2O_1O_3$, where $O_2$ and $O_3$ denote arbitrary orthogonal matrices.
The joint distribution of eigenvalues of this ensemble
$P_{ort}(\varphi_1,\dots,\varphi_N)$ can be found in the book of Girko
\cite{Gi90} and is recalled in the appendix A. In this appendix we
propose a method of generating such matrices numerically and study some
properties of their spectra.

\section{Spectra of products of matrices of circular ensembles}

\subsection {Notation}

We are interested in the spectral properties of products of unitary
matrices, each pertaining to a given ensemble. Let us introduce a
following notation: $D$ denotes a diagonal unitary matrix of CPE, $S$
denotes a symmetric matrix of COE, $U$ represents a unitary matrix of
CUE and $O$ an orthogonal matrix typical to HOE. As usual, the symbol
$SU$ represents a product of two concrete matrices $S$ and $U$. On the
other hand $S*U$ will denote the composed ensemble of unitary matrices
defined as the image of the mapping
\begin{equation}
S_N\times U_N\ni(S,U)\mapsto SU\in U_N
\end{equation}
with the measure induced by this mapping in its image by the product measure
$\mu_S\times\mu_U$ on the Cartesian product $S_N\times U_N$. Indices can be
added to any matrix, if needed. For example $S_1S_2$ denotes a product of two
symmetric matrices, while $S_1*S_2$ represents the composed ensemble defined
as the image of $S_N\times S_N$, which is different from the ensemble of
squared symmetric matrices $S_1*S_1$ obtained by the mapping
\begin{equation}
S_N\ni S\mapsto S^2\in U_N.
\end{equation}

\subsection {Results}

\begin{table}
\begin{tabular}{|c||c|c|c||c|}
\hline
 No &  Composed ensemble &~Measure~&~Spectrum~&~Remarks~  \\
\hline
\hline
A1 &  $U$           &     $\mu_U$         & $P_2$     & CUE    \\
A2 &  $U*S$         &     $\mu_U$         & $P_2$     &  a)    \\
A3 &  $U*D$         &     $\mu_U$         & $P_2$     &  a)    \\
A4 &  $U*O$         &     $\mu_U$         & $P_2$     &  a)    \\
A5 &  $U_1*U_2$     &     $\mu_U$         & $P_2$     & CUE    \\
A6 &  $X_1*U*X_2$   &     $\mu_U$         & $P_2$     &  b)    \\
\hline
B1 &  $S$           &     $\mu_S$         & $P_1$     & COE    \\
B2 &  $U^T*U$       &     $\mu_S$         & $P_1$     &  d)    \\
B3 &  $U^T*D*U$     &     $\mu_S$         & $P_1$     &  d)    \\
B4 &  $S*D$         &     $\mu_S$         & $P_1$     &  g)    \\
B5 &  $S_1*S_2$     &       ?             & $P_1$     &  f)    \\
B6 &  $S_1*S_2*S_1$ &       ?             & $P_1$     &  e)    \\
B7 &  $S_1^{\alpha}*S_2*S_1^{\alpha}$ & ? & $P_1$     &  e)    \\
B8 &  $X^T*S*X$     &       ?             & $P_1$     &  e)    \\
\hline
C1 &  $S_1*S_2*D$   &       ?             & $P_2$     & $n_1$) \\
C2 &  $S_1*S_2*S_3$ &       ?             & $P_2$     & $n_1$) \\
C3 &  $S_1*S_1$     &       ?             &   -       &  h)    \\
C4 &  $S_1*S_1*D$   &       ?             & $P_1$     & $n_2$) \\
C5 &  $S_1*D*S_1$   &       ?             & $P_1$     & $n_2$) \\
\hline
D1 &  $D_1*D_2$     &    $\mu_D$          & $P_0$     &  c)    \\
D2 &  $O_1*O_2$     &    $\mu_O$          & $P_{ort}$ &  c)    \\
D3 &  $O*S$         &       ?             & $P_2$     & $n_3$) \\
D4 &  $O*S*D$       &       ?             & $P_2$     & $n_3$) \\
D5 &  $O_1*D*O_2$   &       ?             & $P_2$     & $n_4$) \\
D6 &  $O*S_1*O^T*S_2$&      ?             & $P_1$     & e), f) (see B8,B5)\\
D7 &  $D_1*O*D_2*O^T$ &     ?             & $P_1$     & $n_5$) \\
D8 &  $O*D_1*O^T*D_1$ &     ?             & $P_1$     & $n_6$) \\
D9 &  $D_1*O_1*D_2*O_1^T*O_2*D_3*O_2^T$&? & $P_2$     & $n_7$) \\
D10 &  $U*D_1*U^T*D_2$  &   ?             & $P_1$     & d), g) (see B3,B4)\\
D11 & $U*D_1*U^{\dagger}*D_2$  & ?        & $P_2$     & $n_8$) \\
D12 & $S*D_1*S^{\dagger}*D_2$  & ?        & $P_2$     & $n_9$) \\
\hline
\end{tabular}
\caption{ Composed ensembles,
 their measures (? represents an unknown measure),  and their
joint probability distribution of eigenvalues. Apart of
ensembles defined in the text, symbol $X$ represents an arbitrary
ensemble of unitary matrices
and $\alpha$ denotes an arbitrary positive real number.}
\end{table}
Main results of this paper concerning the spectra of products of unitary
matrices are collected in Table 1. For convenience we added also some
previously known results. JPD $P_{\beta}$ represents the formula
(\ref{joint}), which depending on $\beta$ describes properties of all
canonical ensembles. Last column of the table gives a reference to the
further text. Some items have not been proved rigorously yet, but are based
on numerical results.

We shall start the discussion of above results with an important note.
The fact that the joint probability distribution of eigenvalues
characteristic to  a given composed ensemble is same as, for example,
for CUE, does not mean at all that the measures of both ensemble are the
same. In other words, if probability measures of two ensembles are equal
$(\mu_a=\mu_b$), then the corresponding JPD are the same $(P_a=P_b)$.
Reverse is not true,
what explains why composition of ensembles is not transitive. For
example JPD of $S$ is the same as for $S_1*S_2$ but differs from this
for $S_1*S_2*S_3$.

\subsection {Remarks and references}

Detailed remarks and references to the table are collected below.

{\bf a)} Let us consider an arbitrary subset $X$ of the unitary group $U_N$
with an arbitrary measure $\mu_X$, and the mapping:
\begin{equation}
f:U_N\times X\ni(U,A)\mapsto UA\in U_N
\label{f}
\end{equation}
The product measure $\mu_{U\times X}=\mu_U\times\mu_X$ in $U_N\times X$
induces a measure in the image of $f$ i.e.\ in $U_N$. Since $\mu_U$ is
left-invariant i.e.\ invariant with respect to the left multiplication by
$V\in U_N$ the same is true for the product measure i.e.\ $\mu_U\times\mu_X$
is invariant under the transformation $(U,A)\mapsto(VU,A)$. In consequence
also the measure induced on $U_N$ is left-invariant. There is only one
(normalized) left-invariant measure on $U_N$ - the Haar measure, hence the
resulting ensemble $U*A$ is CUE. The cases (A2-A5) from the table are
particular examples.

{\bf b)} Since the Haar measure on $U_N$ is also right-invariant an analogous
reasoning shows that $B*U$ gives the CUE ensemble for an arbitrary ensemble
of unitary matrices $X$ from which the matrices $B$ are drawn.  Further,
since $U*A$ and $B*U$ are CUE so is $B*U*A$ for $A$ and $B$ from arbitrary
ensembles $X_1$ and $X_2$ of unitary matrices (The case A6 from the
table above).

{\bf c)} Similar results are valid for diagonal (or orthogonal) matrices.
We must only substitute in the previous reasoning, CUE by CPE (or the
ensemble of orthogonal matrices) with measures $\mu_D$ (or $\mu_O$) and $X$
by an arbitrary subset of diagonal (or orthogonal) matrices. The cases D1
and D2 from the table correspond to this situation.

{\bf d)} It is easy to prove \cite{ZK94} that the mapping
\begin{equation}
g:U_N\ni U\mapsto U^TU\in S_N
\label{g}
\end{equation}
induces in its image (the full set of symmetric unitary matrices) the COE
measure $\mu_S$ i.e.\ in our notation $U^T*U=$COE. This corresponds to the
B2 and B3 in the table.  In the latter case let's observe that
$U^TDU=V^TV$, where $V=D^{1/2}U$ and $D^{1/2}$ denotes an arbitrary diagonal
unitary matrix such that $D^{1/2}D^{1/2}=D$. The mapping
\begin{equation}
U_N\times D_N\mapsto V=D^{1/2}U\in U_N
\end{equation}
induces, according to b), $\mu_U$ in $U_N$ which reduces B3 to B2 with $U$
substituted by $V$.

{\bf e)} Let, as previously, $X$ denote an arbitrary subset of $U_N$ with an
arbitrary probabilistic measure $\mu_X$. From a) it is now clear that the
composite mapping $g\circ f$ where $f$ and $g$ are given by (\ref{f}) and
(\ref{g})
\begin{equation}
g\circ f:U_N\times X\ni (U,A)\mapsto A^TU^TUA\in S_N
\end{equation}
induces COE measure $\mu_S$ in the image $S_N$. Consider now two
following mappings
\begin{eqnarray}
h&:&U_N\times X\ni (U,A)\mapsto (U^TU,A)\in S_N\times X \nonumber
\\
k&:&S_N\times X\ni (S,A)\mapsto A^TSA\in S_N.
\label{hk}
\end{eqnarray}
According to the above, $h$ induces in its image the measure
$\mu_S\times\mu_X$. Since $g\circ f=k\circ h$ they induce the same
measure in their image $S_N$ and, as a consequence, $k$ induces $\mu_S$
in $S_N$ i.e.\ in our notation $A^T*S*A=$COE for $A$ from an arbitrary
ensemble $X$. This corresponds to the case B8 in the table and its
special forms B6 and B7.

{\bf f)} Until now we considered the situations where the ensemble obtained by
multiplication of matrices coincided with CUE, COE or CPE. Our main interest
consists, however, in examination of statistical properties of spectra of
resulting matrices. This allows us to investigate more general situations
in which products either do not have specific symmetry properties or the
induced measure is not equal to $\mu_U$, $\mu_O$, $\mu_S$ or $\mu_D$. As an
example let us consider the mapping
\begin{equation}
s:S_N\times S_N\ni (S_1,S_2)\mapsto S_1S_2\in U_N.
\end{equation}
Observe that the image of $s$ is the whole set $U_N$. Indeed, it is enough
to show that an arbitrary unitary matrix $U$ is a product of two symmetric
unitary matrices. To this end lets denote by $W$ an arbitrary unitary matrix
which diagonalizes $U$ (such a matrix $W$ exists since $U$ is unitary) i.e.
\begin{equation}
U=WDW^\dagger, \quad WW^\dagger=W^*W^T=I
\end{equation}
where $D$ is diagonal and unitary. Now take $S_1=WW^T$ and
$S_2=W^*DW^\dagger$.  Both $S_1$ and $S_2$ are unitary and symmetric and
$S_1S_2=WDW^\dagger=U$.  Nevertheless the measure induced on $U_N$ by
the COE measures $\mu_S\times\mu_S$ on $S_N\times S_N$ is not equal to
CUE measure $\mu_U$. Indeed, for all $S_1,S_2$ the matrix $S_1S_2$ is
unitary similar to $S_1^{1/2}S_2S_1^{1/2}$, where $S_1^{1/2}$ is an
arbitrary unitary, symmetric matrix such that $S_1^{1/2}S_1^{1/2}=S_1$,
(such a unitary, symmetric $S_1^{1/2}$ exists since $S_1$ is unitary and
symmetric). It means that the spectra of $S_1S_2$ and
$S_1^{1/2}S_2S_1^{1/2}=(S_1^{1/2})^TS_2 S_1^{1/2}$ coincide. But from e)
above
we know that the mapping
\begin{equation}
S_N\times X\ni (S_2,S_1^{1/2})\mapsto (S_1^{1/2})^TS_2S_1^{1/2}\in S_N
\end{equation}
induces COE measure $\mu_S$ in the image $S_N$ for $S_2$ from COE and
arbitrary $X$.  It follows that the eigenvalues of
$(S_1^{1/2})^TS_2S_1^{1/2}=S_1S_2$ are distributed according to
(\ref{joint}) with
$\beta=1$, which, on one side, proves that the the mapping does not give CUE
and, on the other side, covers the case B5 from the table.

{\bf g)} Similar reasoning proves the validity of B4. Indeed, observe that
since $S=U^TU$ for some unitary $U$ the matrix $SD=U^TUD$ is unitary similar to
to $UDU^T$, but from already proven case B3 from the table we know that such
multiplication produces COE.

{\bf h)} A superposition of two spectra has JPD different from
canonical
$P_{\beta}$. The two level correlations can be expressed as combination of
correlations of both initial spectra (with rescaled argument) \cite{Dys3},
while level spacing distribution
 may be obtained as a special case of
Berry--Robnik distribution \cite{BR84} (for two equal chaotic layers).

{\bf $n_i$)} Conjectures based on numerical results. Conjectures indexed by
the same index are equivalent. Random orthogonal matrices
where generated as described in Appendix A.
 A modified version of an algorithm  for generation of random unitary
matrices, first presented in Ref.
\cite{ZK94}, is given in the appendix B.
 We generated several realizations of discussed products,
diagonalized them numerically and compared the level spacing distribution
$P(s)$ and number variance $\Sigma^2(L)$ with known predictions of canonical
ensembles \cite{mehta}. Our numerical results are valid thus in the limit of
large $N$ (practically $N\approx20$ and larger). We have performed
additional cross-checking by repeating calculations (with similar results)
using random matrices generated out of eigenvectors.
In order to verify or reject hypothesis concerning
properties of the spectra the long range correlations where found to be more
informative than spacing distribution. In Fig.\ 1 we display number variance
averaged over spectra of exemplary composed ensembles -
 ($O*S, S_1*S_2*D, U*D*U^{\dagger}*D$)
 typical of CUE, and other ($D_1*O*D_1*O^{T}$,
  $D_1*O*D_2*O^{T}$) typical of COE.

 \begin{figure}
\unitlength 1cm
\begin{picture}(9,8)
\put(0.7,10.5){\includegraphics{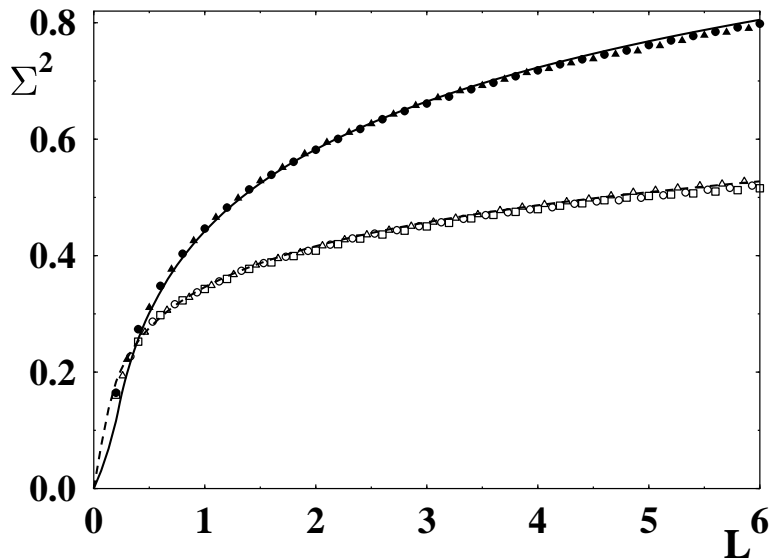}}
 \end{picture}
\unitlength 1bp
\caption{Number variance $\Sigma^2(L)$
 for three ensembles $ D3 (\Box), \
  C1 (\triangle)$,  and $D11 (\circ)$
  with JPD $P_2$ typical of CUE,  and two ensembles
  $ D7 $ and $D8$ (full symbols) with COE like JPD $P_1$.
 Solid and dashed lines stand for  RMT results for COE and CUE.}
\label{fig1}
\end{figure}

{\bf $m$)} Consider composed ensemble defined as a product of $n$
matrices, each pertaining to a given ensemble. For large $n$ we expect
the product to be distributed uniformly with respect to the Haar
measure, thus displaying the CUE-like spectral fluctuations. This remark
obviously hold if at least one matrix belongs to CUE (see ensemble A6).
On the other hand, it does not hold if all $n$ matrices belong to the
Poissonian ensemble, since their product displays the JPD $P_0$.

\subsection {Intermediate ensembles}

Observe that the JPD of the composed ensembles
A5, B5 and D1 can be written as
\begin{equation}
 P[U_{\beta}*U_{\beta}] = P[U_{\beta'}]
\label{betaa}
\end{equation}
with $\beta=\beta'$ equal to $2,1$, and $0$. The number $\beta$,
characterizing the degree of the level repulsion, (\ref{joint}), for
ensembles interpolating between  CPE and CUE may take any real value in
$[0,2]$.

 \begin{figure}
\unitlength 1cm
\begin{picture}(9,8)
\put(0.7,10.5){\includegraphics{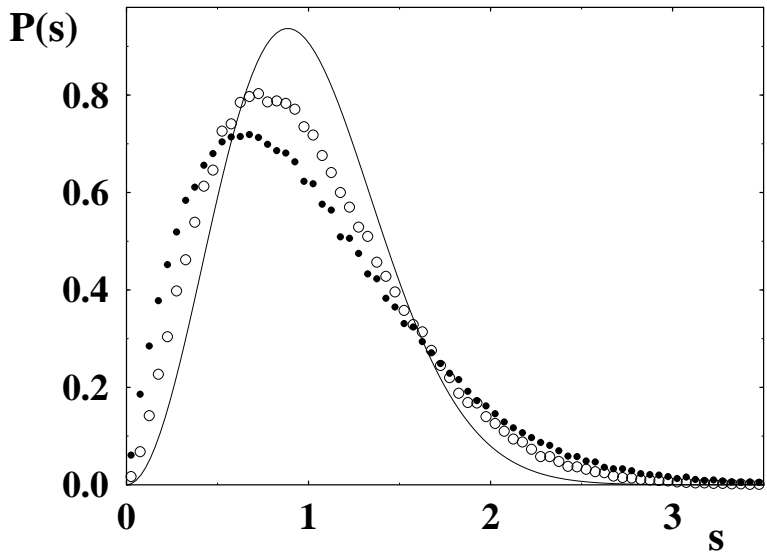}}
 \end{picture}
\unitlength 1bp
\caption{Level spacing distribution  $P(s)$
 for an ensemble $U_{\delta}$  interpolating between Poisson--CUE
  $(\bullet)$, and the composed ensemble $U_{\delta}*U_{\delta}$
$(\circ)$ for the transition parameter $\delta=0.5$.
Solid line represents CUE distribution.
}
\label{fig2}
\end{figure}

In order to investigate,  to what extend the formula (\ref{betaa}) may
be  generalized, we constructed numerically random unitary matrices
pertaining to ensembles interpolating between CPE and CUE as described
in appendix B. Fig.~2 presents level spacing distribution taken of 3000
matrices of size $N=50$, while the value of the parameter $\delta$,
controlling the transition CPE-CUE, is set to $0.5$. Level spacing
distribution of the composed ensemble defined {\it via} product of such
two independent matrices is represented by open symbols. It is closer to
the CUE prediction and can be approximated by distribution typical of
another ensemble with larger value of the control parameter $\delta$. In
other words, for this family of interpolating ensembles the relation
(\ref{betaa}) seems to hold with $\beta'$ being an unknown function of
$\beta$ satisfying $\beta'\ge \beta$.

 \begin{figure}
\unitlength 1cm
\begin{picture}(9,8)
\put(0.7,10.5){\includegraphics{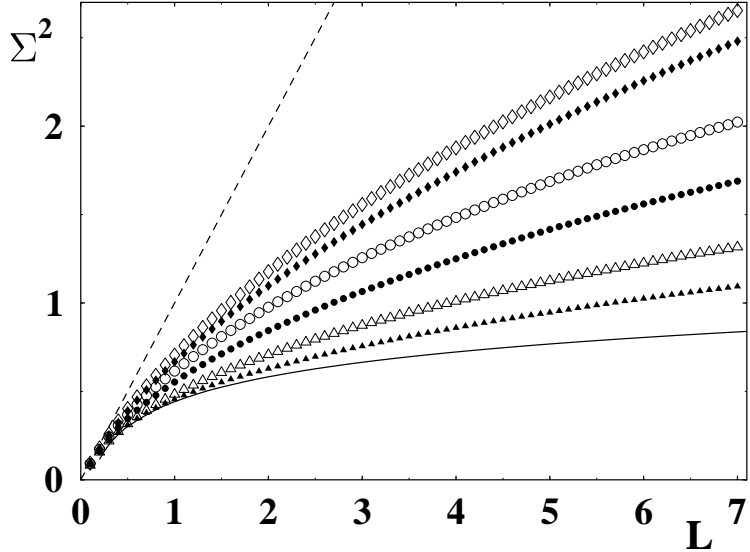}}
 \end{picture}
\unitlength 1bp
\caption{Number variance $\Sigma^2(L)$ for interpolating ensembles
$U_{\delta}$  (open symbols) and the corresponding composed ensembles
$U_{\delta}*U_{\delta}$ (closed symbols) for $\delta=0.1$ $(\Diamond)$,
$\delta=0.3$ $(\circ)$ and $\delta=0.7$ $(\nabla)$. Dashed and solid
lines stand for Poisson and CUE results, respectively.
   }
\label{fig3}
\end{figure}

Further tests of the long range correlations of the spectra allowed us
to support this conjecture. Figure 3 shows the number variance
$\Sigma^2(L)$ \cite{mehta} for simple and composed interpolating
ensembles for three values of the control parameter. In every case the
spectra of products of two matrices (full symbols) are less rigid that
the spectra of the simple interpolating ensemble (open symbols). This
property can be understood realizing that such interpolating unitary
matrices enjoy band structure, as demonstrated in Fig.~4 for an
exemplary matrix of size $N=35$. Vaguely
speaking, a product of two band matrices possess a band of a double
width, and the spectral properties of composed ensembles are thus closer
to these typical of CUE.

 \begin{figure}
\unitlength 1cm
\begin{picture}(9,9)
\put(2,-3.0){\includegraphics{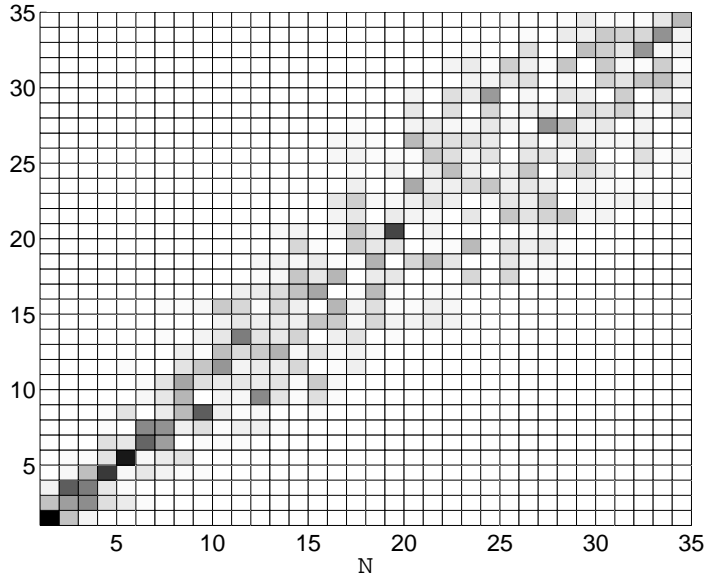}}
 \end{picture}
\unitlength 1bp
\caption{ Squared moduli of elements of a random matrix $|U_{kl}|^2$
taken from
an interpolating ensemble $U_{\delta}$ with $\delta=0.5$. Observe a band
structure of the unitary matrix $U$.
   }
\label{fig4}
\end{figure}

\subsection {Physical applications}

Describing quantized physical systems one encounters often a structure
of one of the above mentioned composed ensembles. Analyzing a concrete
physical system we deal with deterministic matrices, so the assumptions
concerning randomness of each matrix forming the composed ensemble can
not be rigorously fulfilled. It seems however, that the assumptions
concerning randomness are too strong: we provide below examples of
quantum systems which are characterized by JPD found for an appropriate
composed ensemble, although some composing matrices do not display
required properties of presupposed canonical ensembles. To show this one
may study the statistical properties of a {\sl semiclassical ensemble},
i.e.\ the properties of several quantum realizations of the same
classical system, distinguished only by different values of the relative
Planck constant (spin length).

Let us start the discussion analyzing periodically time-dependent
quantum systems. Generally speaking, JPD $P_1$ corresponds to fully
chaotic systems with (generalized) time reversal invariance, while
spectrum characterized by $P_2$ provides an evidence that such a symmetry
has been broken. Let us consider the composed ensemble D6.  A single
orthogonal matrices $O$  pertaining to HOE does not often appear alone
in the theory, nevertheless the compositions $O*D_2*O^T$ are crucial for
many important models. Consider an exemplary periodically kicked system
described by a Hamiltonian $H=H_0+k V\sum_n \delta(t-nT)$. Its free
evolution is represented by $U_1=\exp(T H_0)$ and the perturbation term
can be written as $U_2=\exp(ikV)$, where $V$ is a symmetric operator and
$k$ is the perturbation strength. It is natural to represent the system
in the eigenbasis of $H_0$ so the unitary matrix $D_1=\exp(iTH_0)$ is
diagonal. Orthogonal rotation $O$ allows one to change the basis into
eigenbasis of $V$ and obtain eigenvalues of $U_2$. Note that discussed
ensemble $D_1*O*D_2*O^T$ corresponds just to the Floquet operator
\begin{equation}
F=e^{itH}e^{ikV}
\label{sys1}
\end{equation}
of such a system.  We can therefore expect that if both operators $H$
and $V$ sufficiently do not commute (so as to assure that the transition
matrix $O$ is generic in sense of $\mu_O$), than
for generic values of the parameters $t$ and $k$
the operators $\exp(itH)$
and $\exp(ikV)$ are ''relatively random'' \cite{KZ91} and
 the system described by Floquet operator $F$ is chaotic. In fact
the structure (\ref{sys1}) is typical to several models for {quantum
chaos} discussed in the literature (kicked rotator \cite{Iz90}, kicked
top\cite{hks87,haake}, kicked Harper model \cite{harper}).

In ensemble D6 it is assumed that the diagonal matrices $D_i$ are
random. In the simplest chaotic kicked top model $F_1$, defined by the
angular momentum operators $J_x,J_y,J_z$ acting on $2j+1$ dimensional
Hilbert space as: $F_1=\exp(itJ_z)\exp(ikJ_x^2/2j)$ \cite{hks87}, the
diagonal matrix $D_2$ reads $(D_2)_{lm}=\delta_l^m \exp(ikl^2/2j)$. Due
to the factor $l^2$ in the exponent
for a generic value of the parameter $k$
 the diagonal elements of the matrix $D_2$ are pseudorandom
\cite{gpf84} what assures the COE-like spectral fluctuations of the
orthogonal top $F1$.

To observe the $P_1$ JPD of eigenvalues characteristic to the composed
ensemble D6 it is therefore sufficient, if at least one of the matrices
$D_1$ and $D_2$ is pseudorandom. On the other hand, if both diagonal
matrices $(D_i)_{ll}$ have the structure $\exp(ikl)$, the resulting
operator $F_1'$ does not pertain to COE, what corresponds to the
integrability of the kicked top with $V'=J_x$ \cite{haake}.

In order to get a CUE spectrum it is necessary to break the time
reversal symmetry (or any generalized antiunitary symmetry)\cite{haake}.
As follows from example D8 this can be done by adding additional unitary
term generated by a kick - perturbation $\tilde{V}$ not commuting with
$H$ nor with $V$. This scheme corresponds exactly to the so--called
unitary kicked top given by \cite{hks87}
\begin{equation}
F_2=e^{ik_1J_x^2/2j} e^{itJ_z} e^{ik_2J_y^2/2j},
\label{sys2}
\end{equation}
with $k_1\ne k_2$ (and arbitrary order of unitary factors), or CUE
version of kicked rotator \cite{Iz90}.

According to remark f) the systems which can be brought to a symmetric
COE -like structure by a similarity transformation display spectra
described by $P_1$ JPD. Therefore example B5 represented by $S_1S_2\sim
S_1^{1/2}S_2S_1^{1/2}$ leads to COE spectrum, in contrast to example C1:
$S_1S_2D$, for which such a transformation is not possible. In the same
spirit it is sufficient to modify slightly the system (\ref{sys2}) into
$ F_3=e^{ik_1J_x^2/2j} e^{itJ_z} e^{ik_2J_x^2/2j}$, or
$F_4=e^{ikJ_x^2/2j} e^{itJ_z} e^{ikJ_y^2/2j}$, so as it recovers the
generalized antiunitary symmetry and its spectrum pertains to COE.

Any "unitary" top $F_u$, without time-reversal symmetry, may be
artificially made symmetric by adding the same sequences $F_u$ of
perturbation in the reverse order. Therefore $F=F_uF_u^T$ displays COE
like fluctuations of the spectra. Mathematical theory of time reversible
and irreversible tops is given in \cite{hks87}, while some further
examples where numerically studied in \cite{Zy93}.

A product of two symmetric random matrices $S_1S_2$ arises in the theory
of chaotic scattering \cite{smil,gopar,Zy97}. Its spectrum obeys COE
statistics, as follows from the example B5. The same statistics is
characteristic to several versions of quantized Baker map
\cite{BV89,Sa91,fra}, which is also represented by a product   of two
symmetric matrices $B={\cal F}_1 {\cal F}_2$, although both matrices
${\cal F}_1$ and ${\cal  F}_2$, defined via Fourier matrices, do not
show the properties of   COE.

As a last example let us consider the piecewise affine transformation of
the torus, which can be quantized as \cite{fra} $T=D_2{\cal F}^{\dagger}
D_1 {\cal F}$. Diagonal matrices, of the type discussed above,
$(D_1)_{ll}=\exp(i a l^2)$ are pseudorandom for a generic value of the
parameter $a$. Albeit the symmetric Fourier matrix ${\cal F}$ is not
typical to COE, the structure of $T$ resembles the ensemble D12, and its
spectrum confers to the predictions of CUE.

\section{Concluding remarks}
Let us conclude our paper with the following, summarizing remarks.
Various statistical properties of products of random matrices can be
interesting when studying quantum chaotic systems influenced by symmetry
breaking perturbations. We showed that using our results we can predict
properties of spectra of a large class of periodically driven model
systems (kicked tops).

From the mathematical point of view our investigations leave many
questions open. Not in all cases we were able to calculate the resulting
probability distribution of the composite ensembles.  In fact it was
possible only in those cases where the distribution coincided with one
of the "classical" ones (COE, CUE, CPE, HOE). In some cases for which we
did not find the probability distribution of the ensemble we were
nevertheless able to give the corresponding distribution of the
eigenvalues, from which the most popular statistical measure of quantum
chaotic systems, namely the distribution of neighboring levels, is
easily calculable. For some other composed ensembles we
provided numerical evidence for
their distribution of eigenvalues
applying efficient methods of constructing of random
unitary ensembles of all canonical ensembles.
Further investigation should resolve the problem of
the full probability distributions for these composed examples and find
analytical arguments for distributions  of eigenvalues
founded numerically.

\acknowledgments
We acknowledge financial support by Polish Committee of Scientific
Research under the Grant No.~2P03B~03810.

\appendix

\section{Random orthogonal matrices}

The distribution of eigenvalues in the ensemble of random orthogonal
matrices can be found in the books of Girko \cite{Gi90,Gi88}. We shall
give here the relevant result for completeness. The distribution
density of matrices in the ensemble is the (normalized) Haar measure on
the orthogonal group $o(N)$.  The simpler situation occurs for $N$ odd.
In this case among the eigenvalues there is one (say $\phi_0$) equal $1$
or $-1$.  The rest of eigenphases can be grouped into pairs
$(\phi_i,-\phi_i), -\pi<\phi_i<\pi$, $i=1,\ldots,(N-1)/2$.  With the
probability $1$ they are not degenerate and distributed independently of
eigenvectors. The joint probability distribution of eigenphases reads
\begin{eqnarray}
P(\phi_1,\ldots,\phi_{(N-1)/2}, \pm 1)
={\cal N}\prod_{n=1}^{(N-1)/2}
\left((1\pm 1)\sin^2\frac{\phi_n}{2}+(1\mp 1)\cos^2\frac{\phi_n}{2}\right)
|\sin\phi_n|
\prod_{k<n}\sin^2\frac{\phi_n-\phi_k}{2}\sin^2\frac{\phi_n+\phi_k}{2},
\end{eqnarray}
where the last argument $\pm 1$ and the alternative signs in the rest of
the formula refer to $\phi_0=1$ or $\phi_0=-1$. For the slightly more
complicated case of $N$ even consult the above cited books of Girko.

In order to generate numerically a random orthogonal matrix typical of
HOE we employed a parametrisation of the orthogonal group defined by
Hurwitz in the classical paper \cite{hur} published exactly one hundred
years ago. An arbitrary $N$ dimensional orthogonal matrix $O$ may be
written as a product of $N(N-1)/2$ elementary orthogonal rotations in
two-dimensional subspaces. The matrix of such an elementary orthogonal
rotation will be denoted by $F^{(i,j)}(\psi)$.  The only nonzero
elements of $F^{(i,j)}$ are
\begin{eqnarray}
\label{oeel}
F^{(i,j)}_{kk}=1, &\quad & k=1,\dots,N; \quad k\neq i,j \nonumber   \\
F^{(i,j)}_{ii}=\cos\psi, &\quad & F^{(i,j)}_{ij}=\sin\psi,\nonumber \\
F^{(i,j)}_{ji}=-\sin\psi,&\quad & F^{(i,j)}_{jj}=\cos\psi .
\end{eqnarray}
From these transformations one constructs the following $N-1$ composite
orthogonal rotations
\begin{eqnarray}
\label{oecom}
F_1&=&F^{(N-1,N)}(\psi_{01}),
\nonumber \\
F_2&=&F^{(N-2,N-1)}(\psi_{12})F^{(N-1,N)}(\psi_{02}), \nonumber \\
F_3&=&F^{(N-3,N-2)}(\psi_{23})F^{(N-2,N-1)}(\psi_{13})
F^{(N-1,N)}(\psi_{03}), \\
\ldots \nonumber \\
F_{N-1}&=&F^{(1,2)}(\psi_{N-2,N-1})F^{(2,3)}
(\psi_{N-3,N-1})\ldots
 F^{(N-1,N)}(\psi_{0,N-1}), \nonumber
\end{eqnarray}
and finally forms the orthogonal transformation $O$ as
\begin{equation}
O= F_1F_2F_3 \ldots F_{N-1}.
\label{ouhur}
\end{equation}
Uniform distribution with respect to the Haar measure on the orthogonal
group is achieved if the generalized Euler angles $\psi_{0s}$ are
uniformly distributed in the interval
 $0\leq\psi_{0s}< 2\pi$, and the remaining angles $\psi_{rs}$ (for
$r>0$) are taken from the interval $[0, \pi]$ according to the measure
$d\mu_r=(\sin \psi_{rs})^r d\psi_{rs}$ \cite{hur}. An alternative way to
generate random orthogonal matrices was recently proposed by Heiss
\cite{Ha94}. Random orthogonal matrices may also be obtained as
eigenvectors of real random symmetric matrices typical to Gaussian
orthogonal ensemble.

 \begin{figure}
\unitlength 1cm
\begin{picture}(9,8)
\put(0.7,10.5){\includegraphics{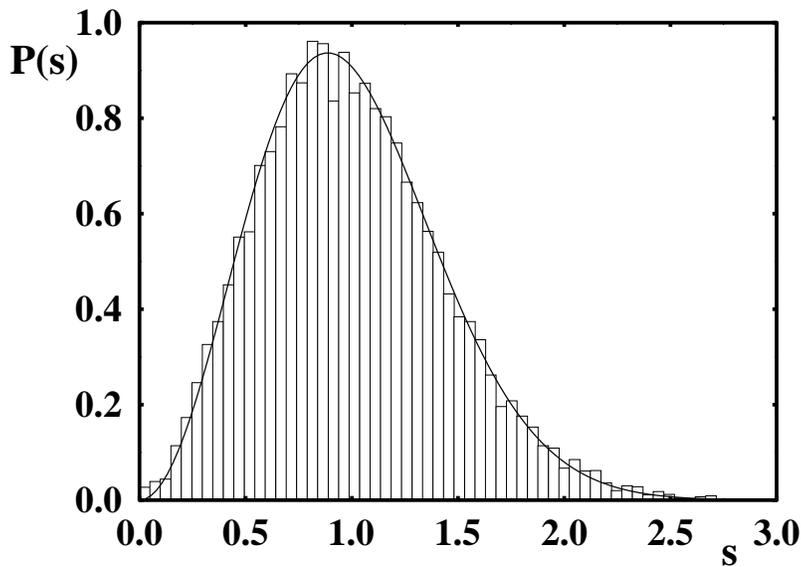}}
 \end{picture}
\unitlength 1bp
\caption{Level spacing distribution  $P(s)$
 for an ensemble of orthogonal matrices
 distributed uniformly with respect to the Haar measure  (histogram)
 may be
 approximated by the CUE formula (solid line).
}
\label{fig5}
\end{figure}

Figure 5 presents the level spacing distribution obtained of 20000
random orthogonal matrices obtained by Hurwitz parametrisation for
$N=41$. Numerical data suggest that statistical properties of the
spectra of random orthogonal matrices are close to the CUE predictions.

\section{Random unitary matrices}

In our earlier paper \cite{ZK94} we have also used the Hurwitz
\cite{hur} parametrisation to generate random unitary matrices. We
present it here in details for completeness of the present paper and
since in the text of \cite{ZK94} a slightly different, not yet verified
algorithm appeared (the numerical calculation, however, were based on
the prescription given below).

An arbitrary unitary transformation $U$ can be composed from elementary
unitary transformations in two-dimensional subspaces. The matrix of such
an elementary unitary transformation will be denoted by
$E^{(i,j)}(\phi,\psi,\chi)$.  The only nonzero elements of $E^{(i,j)}$
are
\begin{eqnarray}
\label{eel}
E^{(i,j)}_{k k}&=&1,\quad k=1,\dots,N; \quad k\neq i,j \nonumber \\
E^{(i,j)}_{i i}&=&\cos\phi e^{i\psi},\nonumber \\
E^{(i,j)}_{i j}&=&\sin\phi e^{i\chi},  \\
E^{(i,j)}_{j i}&=&-\sin\phi e^{-i\chi},\nonumber \\
E^{(i,j)}_{j j}&=&\cos\phi e^{-i\psi}.\nonumber
\end{eqnarray}
From the above elementary unitary transformations one constructs the
following $N-1$ composite rotations
\begin{eqnarray}
\label{ecom}
E_1&=&E^{(N-1,N)}(\phi_{01},\psi_{01},\chi_{1}),
\nonumber \\
E_2&=&E^{(N-2,N-1)}(\phi_{12},\psi_{12},0)E^{(N-1,N)}(\phi_{02},\psi_{02},
\chi_{2}), \nonumber \\
E_3&=&E^{(N-3,N-2)}(\phi_{23},\psi_{23},0)E^{(N-2,N-1)}(\phi_{13},\psi_{13},
0)E^{(N-1,N)}(\phi_{03},\psi_{03},\chi_{3}), \\
\ldots \nonumber \\
E_{N-1}&=&E^{(1,2)}(\phi_{N-2,N-1},\psi_{N-2,N-1},0)E^{(2,3)}
(\phi_{N-3,N-1},\psi_{N-3,N-1},0)\ldots
\nonumber \\
&\ldots&E^{(N-1,N)}(\phi_{0,N-1},\psi_{0,N-1},\chi_{N-1}), \nonumber
\end{eqnarray}
and finally forms the unitary transformation $U$ as
\begin{equation}
U=e^{i\alpha}E_1E_2E_3 \ldots E_{N-1}.
\label{uhur}
\end{equation}
The angles $\alpha, \phi_{rs},\psi_{rs}$, and $\chi_{s}$ are
taken uniformly from the intervals
\begin{equation}
 0\leq\psi_{rs}<2\pi \delta, \quad
0\leq\chi_{s}<2\pi\delta, \quad 0\leq\alpha<2\pi \delta,
\label{intv}
\end{equation}
whereas
\begin{equation}
\phi_{rs}={\arcsin}(\xi_{rs}^{1/(2r+2)}),\quad r=0,1,2,\ldots,N-2
\label{phi}
\end{equation}
with $\xi_{rs}$ uniformly distributed in
\begin{equation}
0\leq\xi_{rs}<\delta, \quad \quad 0\leq r < s \leq N-1.
\label{xi}
\end{equation}
If the parameter $\delta$ is set
to unity
then the obtained matrix is drawn from the Circular Unitary Ensemble
\cite{hur}.

In order to obtain a a family of ensembles interpolating between diagonal
matrices of CPE and generic unitary matrix typical of CUE we construct a
product $U_{\delta}=D{\hat {U}_{\delta}}$. Diagonal matrix $D$ is
typical of CPE, while the matrix ${\hat{U}}_{\delta}$ is obtained
according the above procedure with real parameter $\delta\in (0,1)$
determining the intervals in Eq. (\ref{intv}) and (\ref{xi}). Varying
the value of this parameter from zero to unity one obtains a continuous
interpolation between CPE and CUE \cite{ZK96}.

Random unitary matrices may be also constructed by taking $N$
eigenvectors of random Hermitian matrix pertaining to the Gaussian
unitary ensemble. In this procedure one must specify $N$ arbitrary
phases of each eigenvector. This method, albeit simple, does not allow
to control parameters of the interpolating ensemble as it is possible for the
Hurwitz algorithm discussed above.

\end{document}